# Monitoring Control Updating Period In Fast Gradient Based NMPC

Mazen Alamir[1]

*Abstract*— In this paper, a method is proposed for on-line monitoring of the control updating period in fast-gradient-based Model Predictive Control (MPC) schemes. Such schemes are currently under intense investigation as a way to accommodate for real-time requirements when dealing with systems showing fast dynamics. The method needs cheap computations that use the algorithm on-line behavior in order to recover the optimal updating period in terms of cost function decrease. A simple example of constrained triple integrator is used to illustrate the proposed method and to assess its efficiency.

## I. INTRODUCTION

A great amount of research during the last decade has been dedicated to the implementation of MPC schemes to systems showing fast dynamic. These are systems having characteristic time requiring updating periods that are incompatible with the complete solution of the underlying optimization problem.

The solution that emerges with a wide consensus is to perform, during each control updating period, only a limited number of iterations of some descent method and to apply the so obtained sub-optimal solution during the next updating period. The iterations are then continued after a potential warm start at the next updating period leading hopefully to asymptotical recovering the optimal solution.

In [2], the author showed that this implementation framework enhances a generic trade-off in terms of the control updating period. Indeed, if the latter is too small, the decrease in the cost function is not sufficient to compensate for the increase that may be due to model uncertainties (including those related to the potentially unknown dynamic of the set-point to be tracked). On the other hand, For too large control updating periods, the lack of fresh information means that the optimizer would solve with a high precision a rather out-of-date optimization problem.

In the present paper, the generic framework proposed in [2] is adapted to the specific context where the iterations are defined by gradient or fast gradient algorithms (see for instance [3], [4], [8] and the reference therein). More precisely, while in [2], attempt to identify the so-called efficiency map that characterizes the descent method is considered, here the number of iterations (and hence the control updating period) is adapted locally based on the estimation of the gradient of the settling time under a contraction constraint on the dynamic of the cost function.

This method can be applied as such when pure gradient descent method is applied to distribute the computation of the optimal control sequence. In the case where a fast gradient iterations are used, it is shown that the restarting mechanism proposed by [7] is necessary in order to obtain a monotonically decreasing cost function.

The paper is organized as follows. First, the general framework proposed in [2] is recalled in section II in order to set the notation and to underline the crucial role played by the so called iteration efficiency map involved in a small gain-like theorem. An ideal optimization problem is then derived in which the decision variable is the control updating period (or equivalently, the number of iterations before updating the control) while the cost function represents the settling time of the MPC closed-loop cost function evolution. In Section III, an updating rule for the control updating period is defined towards the minimization of this cost function and is based on the estimation of its gradient. This estimation is done using available data regarding the behavior of the algorithm. For the proposed updating rule to asymptotically solve the ideal optimization problem, a monotonicity property is needed for the descent method. Section IV recalls the fast gradient method and shows how restarting mechanism [7] can be used to enhance monotonic decrease of the fast gradient method making the proposed updating rule appropriate. Finally, in section V, an illustrative example is given in order to show the efficiency of the proposed scheme in monitoring on-line the number of iterations of the fast gradient algorithm when used in real-time implementation of constrained MPC. Finally, section VI concludes the paper and gives hints for further investigation.

## II. THE IDEAL CONTROL UPDATING RULE

### A. Some definitions and notation

In the present paper the framework of parameterized NMPC [1] is adopted for the sake of generality since the standard piece-wise constant control parameterization can be viewed as a particular instantiation of the following more general form:

$$\mathcal{U}_{pwc}(p) := \bigl(u^{(1)}(p),\dots,u^{(N)}(p)\bigr) \in \mathbb{U} \subset \mathbb{R}^{N \cdot n_u} \quad (1)$$

where $p \in \mathbb{R}^{n_p}$ is the vector of degrees of freedom while the $u^{(i)}(p)$ stands for the corresponding constant value of the control vector at the future $i$-th sampling period.

Using the notation above, the model of the dynamic system can be given in the following general form:

$$\forall j \in \{1,\dots,N\} \quad , \quad x(k+j) = X(j, x(k), p) \quad (2)$$

where $X(j, x, p)$ is the state reached by the model at the $j$-th future sampling instant when it starts at the state $x(k)$ under the p.w.c sequence defined by $p$ through (1).

[1]CNRS-University of Grenoble, Control Systems Department 11 rue des mathématique, Domaine Universitaire, Saint Martin d'Hères, France. `mazen.alamir at grenoble-inp.fr`

Since the difference between the model used by the controller and the *real* system plays a crucial role in the sequel, the notation:

$$X^r(j, x(k), p, \mathbf{w})$$

is used hereafter to denote the state of the real system after $j$ sampling periods, starting from $x(k)$, under the control sequence defined by $p$ and when some disturbance sequence $\mathbf{w}$ occurs. The shorter notation $X^r(j, x(k), p)$ is sometimes used (explicit mention of $\mathbf{w}$ is omitted) as no specific discussion is needed regarding the origin of $\mathbf{w}$.

Let us denote by $J(p, x)$ the value of the cost function that is used to define a Model Predictive Controller (MPC). The following assumption is considered in the sequel which is always possible to meet by adding appropriate constant to the cost function

**Assumption 1:** There is some $\underline{J} > 0$ such that

$$\forall (p, x), \quad J(p, x) \geq \underline{J} > 0 \qquad (3)$$

This assumption is needed since contraction ratios of the cost function at successive instants are used in the sequel in which the cost function may appear in the denominator.

Obviously, such a cost function is defined for a given control sequence parameter $p$ and for a given current state $x$ since these quantities define the state and control trajectories of the system's model. The prediction horizon $N$ is implicitly included in the expression of $J$ as it is considered to be given once for all in the current paper.

An ideal MPC is defined by $u^{(1)}(\hat{p}(x))$ [see (1)] where $\hat{p}(x)$ is the solution of the following constrained optimization problem:

$$\hat{p}(x) := \min_{p \in \mathbb{R}^{n_p}} J(p, x) \quad \text{under } C(p, x) \leq 0 \qquad (4)$$

where $C(p, x)$ defines the problem constraints.

*B. Distributed-in-time Implementation of MPC*

Nowadays real-time MPC related investigations are justified by the fact that the computation of $\hat{p}(x)$ would be impossible during a single sampling period. Rather a limited number $q$ of iterations of some descent method are performed. This can be denoted by:

$$p^{(q)} = \mathcal{S}^q(p^{(0)}, x) \qquad (5)$$

where $p^{(q)}$ is the parameter vector that is obtained after $q$ iterations of some descent method[1] $\mathcal{S}$ starting from some initial value $p^{(0)}$. The map is obviously parameterized by $x$ since the optimization problem (4) is also parameterized by the current state.

**Assumption 2:** A single iteration of the subroutine $\mathcal{S}$ needs a fixed amount of computation time that is denoted hereafter by $\tau_c$.

**Assumption 3:** The period $\tau_c$ is used as the basic sampling period for the definition of the p.w.c control profile given by (1). Namely each value $u^{(i)}(p)$ is maintained constant during $\tau_c$ time units.

---
[1]The gradient and fast gradient iteration are considered in the present contribution.

Therefore, if several iterations $q$ are needed, in a sense that will be clearer in the sequel, the control parameter $p$ cannot be updated before $\tau_u = q\tau_c$ time units. $\tau_u$ may be called the control updating period. However, if the number of iterations $q$ varies dynamically (as it is proposed in the current contribution) then it is no more possible to define a control updating *period* but only control updating instants $t_k^u$ such that:

$$t_{k+1}^u = t_k^u + q(t_k^u) \cdot \tau_c \qquad (6)$$

where $q(t_k^u)$ is the number of iterations to be performed before the next updated value of the control parameter $p(t_{k+1}^u)$ is delivered. Meanwhile, the previously computed control sequence:

$$u^{(1)}(p(t_k^u)), \ldots, u^{(q(t_k^u))}(p(t_k^u)) \qquad (7)$$

is applied during the updating interval $[t_k^u, t_{k+1}^u]$. In the sequel, the following notation is used:

$$\tau_k^u = t_{k+1}^u - t_k^u = q(t_k^u) \cdot \tau_c \qquad (8)$$

Note that for this scheme to be possible, the following constraint is obviously needed:

$$q(t_k^u) \leq N \quad (N = \text{is the prediction horizon}) \qquad (9)$$

since otherwise, the sequence invoked in (7) would not be available.

**To summarize**, the implementation scheme can be described as follows:

1) At initial instant $t_0^u = 0$, some initial parameter vector $p(t_0^u)$ is chosen. An initial number of iterations $q_0 = q(t_0^u) \leq N$ is also decided.
2) During the updating interval $[t_0^u, t_1^u = q_0\tau_c]$ the first $q_0$ elements of the control sequence $\mathcal{U}(p(t_0^u))$ are applied.
3) In parallel, the computation unit performs the following tasks during $[t_0^u, t_1^u]$:
   a) Compute a model based prediction $\hat{x}(t_1^u)$ of the state at instant $t_1^u$.
   b) Compute $p(t_1^u) = \mathcal{S}^{(q(t_0^u))}(p^+(t_0^u), \hat{x}(t_1^u))$ where the initial guess $p^+(t_0^u)$ is either equal to $p(t_0^u)$ (cold start) or equal to some warm start value that is associated to $p(t_0^u)$ by standard translation technique.
4) At the updating instant $t_1^u$, decide the number of iterations $q(t_1^u)$ to perform during the next updating interval $[t_1^u, t_1^u + q(t_1^u)\tau_c]$. The way this choice is done represents the contribution of the paper and is explained in the following sections.
5) During the updating interval $[t_1^u, t_1^u + q(t_1^u)\tau_c]$, apply the first $q(t_1^u)$ elements of the sequence $\mathcal{U}(p(t_u^1))$ previously computed
6) keep doing...

*C. Monitoring Control Updating As A Feedback Problem*

It comes out from the discussion above that an extended system can be defined in which the state is given by $z :=$

$(p^T, x^T)^T$ and for which the dynamics is given by:

$$x(t_{k+1}^u) = X^r(\tau_u^k, x(t_k^u), p(t_k^u), \mathbf{w}) \quad (10)$$
$$p(t_{k+1}^u) = \mathcal{S}^{q(t_k^u)}\big(p^+(t_k^u), \underbrace{X(\tau_k^u, x(t_k^u), p(t_k^u))}_{\hat{x}(t_{k+1}^u)}\big) \quad (11)$$

or in a more condensed way using the extended state $z$:

$$z(t_{k+1}^u) = F(z(t_k^u), q(t_k^u), \mathbf{w}) \quad (12)$$

which can be viewed as a dynamical system in which $z$ plays the role of state vector while $q$ plays the role of control input. Moreover, the control objective that has to be achieved by the control input $q$ is to steer the scalar output $J(p, x) = J(z)$ to its minimum value as fast as possible. Note that whether this achieves the control objective depends on the adequacy of the original MPC formulation and lies out of the scope of the control input $q$. The latter aims simply at making the distributed-in-time optimization reaches asymptotically the behavior of the ideal MPC computation.

In order to analyze the evolution of the cost function, the following notation is needed:

$J_k^+ := J(p^+(t_k^u), \hat{x}(t_{k+1}^u))$: The value of the cost function when the iterations start at the beginning of the updating interval $[t_k^u, t_{k+1}^u]$ based on the predicted value $\hat{x}(t_{k+1}^u)$ of the future state at instant $t_{k+1}^u = t_k^u + q(t_k^u) \cdot \tau_c$.

$\hat{J}_{k+1} := J(p(t_{k+1}^u), \hat{x}(t_{k+1}^u))$: The value of the cost function after $q(t_k^u)$ iterations based on the predicted value of $\hat{x}(t_{k+1}^u)$ of the state.

$J_{k+1} := J(p(t_{k+1}^u), x(t_{k+1}^u))$: The *true* value of the cost function at the new value of the state $x(t_{k+1}^u)$.

Obviously, the convergence is tightly related to the ratio $J_{k+1}/J_k$ which can be written in terms of the above quantities in the following form:

$$\frac{J_{k+1}}{J_k} = \underbrace{\left[\frac{\hat{J}_{k+1}}{J_k^+}\right]}_{E_k(q(t_k^u))} \times \underbrace{\left[\frac{J_{k+1}}{\hat{J}_{k+1}}\right]}_{D_k(q(t_k^u))} \times \left[\frac{J_k^+}{J_k}\right] \quad (13)$$

where

- The term $E_k(q(t_k^u))$ only depends on the efficiency of the iteration scheme that starts at $(p^+(t_k^u), \hat{x}(t_{k+1}^u))$ as it represents the *local* ratio between the value at the beginning of the $q(t_k^u)$ iterations and the value at the end and this for a given estimation $\hat{x}(t_{k+1}^u)$ of the future state.

- The term $D_k(q(t_k^u))$ represents the effects of uncertainties and/or the imperfection of the finite horizon parameterization. More precisely:
  - the term $J_{k+1}/\hat{J}_{k+1}$ is linked to the difference between the predicted state $\hat{x}(t_{k+1})$ that is used in the iterations and the true state $x(t_{k+1})$ that is effectively found at $t_{k+1}^u$. This difference may be induced by model mismatches (including unknown dynamics of the set-point when the latter is used to define an extended state for instance).

  - the term $J_k^+/J_k$ represents the model based ratio between the new value of the cost function after prediction horizon shift by $\tau_k^u$ and using the shifted value of the control parameter $p^+(t_k^u)$ and the value of the cost function before horizon shift. This ratio would be necessarily < 1 in an infinite horizon setting using classical p.w.c control parameterization.

Based on (13) it comes clearly that the indicator $K_k^{min}$ defined by:

$$K_k^{min} := \min_{q \in \{1,\dots,N\}} K_k(q) := E_k(q) D_k(q) \quad (14)$$

is of great importance since the convergence of the scheme may be guaranteed if for all $k$, one can be sure that

$$K_k^{min} < 1$$

The computation of the *feedback control* $q(t_k^u)$ in (12) can be rationally done if one disposes at instant $t_k^u$ of estimated models for the maps $E_k(q)$ and $D_k(q)$ as functions of the number of iterations $q$. Indeed, in this case, the following *feedback* can be used:

$$q^*(t_k^u) := \arg\min_{q \in \{1,\dots,N\}} \Phi(q) := \quad (15)$$

$$\begin{cases} \dfrac{q}{|\log(K_k(q))|} \text{ under } K_k(q) < 1 & \text{If } K_k^{min} < 1 \\ K_k(q) & \text{otherwise} \end{cases}$$

More clearly, if the constraint $K_k(q) < 1$ is feasible ($K_k^{min} < 1$), then optimization focuses on the settling time which is proportional to $q/|\log(K_k(q))|$ but this minimization is done over those values of $q$ that correspond to a contraction. Otherwise the contraction factor is enforced by minimizing $K_k(q)$.

To summarize, if models for $E_k(\cdot)$ and $D_k(\cdot)$ were available, equation (15) completely defines the feedback law $q$ for the extended system (12). It may be argued however that assuming the availability of $E_k(q)$, even through identification (as it has been suggested and shown to be efficient on a rather involved nonlinear example in [2]) is a rather strong assumption. An alternative method is proposed in the next section that enhances a distributed-in-time solution of (15).

III. A SUB-OPTIMAL CONTROL UPDATING RULE

Recall first of all that at the end of the computation period $[t_k^u, t_{k+1}^u]$, one disposes by using (13) of the following local estimation of $E_k(q(t_k^u))$ and $D_k(q(t_k^u))$:

$$E_k(q(t_k^u)) = \frac{\hat{J}_{k+1}}{J_k^+} \quad ; \quad D_k(q(t_k^u)) = \left[\frac{J_{k+1} J_k^+}{\hat{J}_{k+1} J_k}\right] \quad (16)$$

Moreover, provided that $q(t_k^u) \geq 2$, one can use the intermediate results of the iterations to estimate the gradient of $E_k$

at $q(t_k^u)$. Indeed, one obviously has:

$$\frac{\Delta E_k}{\Delta q}(q(t_k^u)) \approx$$
$$\frac{J(p^{(q(t_k^u))}, \hat{x}(t_{k+1}^u)) - J(p^{(q(t_k^u)-1)}, \hat{x}(t_{k+1}^u))}{J(p^{(0)}, \hat{x}(t_{k+1}^u)} \quad (17)$$

Regarding the map $D_k(\cdot)$, one may notice that by definition $D_k(0) = 1$. Therefore, the following simple model can be adopted for $D_k(\cdot)$:

$$D_k(q) = 1 + \alpha_k^D \cdot q \quad (18)$$

where the coefficient $\alpha_k^D$ is estimated, in accordance with (16) by the expression:

$$\alpha_k^D := \frac{1}{q(t_k^u)}\left[\frac{J_{k+1}J_k^+}{\hat{J}_{k+1}J_k} - 1\right] \approx: \frac{\Delta D_k}{\Delta q}(q(t_k^u)) \quad (19)$$

which is nothing but an estimation of the gradient of $D_k$ at $q(t_k^u)$. Therefore, using (17) and (19), the sensitivity of $K_k$ at $q(t_k^u)$ can be evaluated:

$$\frac{\Delta K_k}{\Delta q}(q(t_k^u)) \approx E_k \frac{\Delta D_k}{\Delta q} + D_k \frac{\Delta E_k}{\Delta q} \quad (20)$$

Moreover, when $K_k(q(t_k^u)) < 1$, the sensitivity of the settling time can also be obtained by:

$$\frac{\Delta(q/|\log(K_k(q))|)}{\Delta q} \approx \frac{-\log(K_k) + \frac{q}{K_k} \times \frac{\Delta K_k}{\Delta q}}{[\log(K_k)]^2} \quad (21)$$

Having all the quantities above at hand, the updating rule for $q$ can be given by Algorithm 1:

---

**Algorithm 1** Updating rule $q(t_{k+1}^u) = U(q(t_k^u), t_k^u)$

1: **If** $(K_k \geq 1)$ **then**
2: $\quad \Gamma \leftarrow \frac{\Delta K_k}{\Delta q}(q(t_k^u))$     [see (17), (19) and (20)]
3: **Else**
4: $\quad \Gamma \leftarrow \frac{\Delta(q/|\log(K_k(q))|)}{\Delta q}$     [see (21)]
5: **End If**
6: $q(t_{k+1}^u) \leftarrow \max\{2, \min\{q_{max}, q(t_k^u) - \delta \cdot sign(\Gamma)\}\}$

---

Note that in Algorithm 1, $\Gamma$ stands for the gradient of $K_k$ if the current estimated value of $K_k$ is greater to 1 (Step 2), otherwise, $\Gamma$ is taken to be the gradient of the settling time (Step 4). In both cases, a step $\delta \in \mathbb{N}$ in the opposite direction is taken and resulting value is projected in $[2, q_{max}]$ where $q_{max} \leq N$ is the maximum number of iterations being allowed. Note that $q \geq 2$ is needed for the estimation of $\Delta E_k/\Delta q$ to be possible according to (17).

It is obvious that Algorithm 1 implements a quantified gradient method that hopefully reaches a vicinity of the optimal solution of (15). This is more likely to occur when the unknown efficiency map is monotonically decreasing function of $q$. This is generally not the case in fast gradient method unless a restarting strategy is adopted as shown in the following section.

## IV. MONOTIC VERSION OF THE FAST GRADIENT METHOD

The fast gradient [5] version of the descent iteration invoked in (5) leads to the following algorithm:

---

**Algorithm 2** Fast gradient iterations $\bar{p} = \mathcal{S}^{(q)}(p, x)$

1: Initialization. $p^{(0)} \leftarrow p$, $r \leftarrow p$
2: **for** $i = 1, q$ **do**
3: $\quad p^{(i)} \leftarrow P_C\left(r - \frac{1}{L}[\nabla J(r, x)]\right)$
4: $\quad r \leftarrow p^{(i)} + c(p^{(i)} - p^{(i-1)})$
5: **end for**
6: $\bar{p} \leftarrow p^{(q)}$

---

in which $L$ is an upper bound on the Lypschitz constant of the gradient $\nabla J$ of the cost function, namely:

$$\forall(x, p_1, p_2) \,,\, \|\nabla J(p_2, x) - \nabla J(p_1, x)\| \leq L\|p_2 - p_1\|$$

while $c \in [0, 1[$ is chosen according to the problem conditioning [6]. For quadratic problems, the choice $c = \frac{\sqrt{\lambda_{max}(H)} - \sqrt{\lambda_{min}(H)}}{\sqrt{\lambda_{max}(H)} + \sqrt{\lambda_{min}(H)}}$ where $H$ is the problem Hessian is considered to be optimal. Note that in the case where $c = 0$ is used, the algorithm reduces to the classical pure gradient descent.

The map $P_C$ is the projection map into the set of admissible values. This map can be made easy to compute by working on the dual problem in which the constraints are limited to hypercube for which $P_C$ is reduced to a vector of element-wise saturation constraints (See for instance [3], [4]).

The convergence of the above scheme depends on the choice of $c$. Indeed, the pure gradient method ($c = 0$) shows a proved convergence rate in $1/i$ while an optimal fast gradient method decreases the cost function as $1/(i+2)^2$. More precisely, denoting by $J^*(x)$ the optimal value of the cost function, one can write:

$$J(\mathcal{S}^{(q)}(p, x), x) - J^*(x) \leq \frac{\alpha_0}{q^\ell} \quad ; \quad \ell \in [1, 2] \quad (22)$$

Recall that while the pure gradient method leads to monotonically decreasing cost, the use of $c > 0$ in the fast gradient version very frequently leads to non monotonically decreasing behavior[2]. In [7], the authors showed very elegantly that this fact occurs systematically under certain quite mild conditions. Moreover, they proposed the following modified version of the fast algorithm:

---

[2]It is worth underlying that this is not incompatible with (22) since the later is only a bound on the potential oscillations.

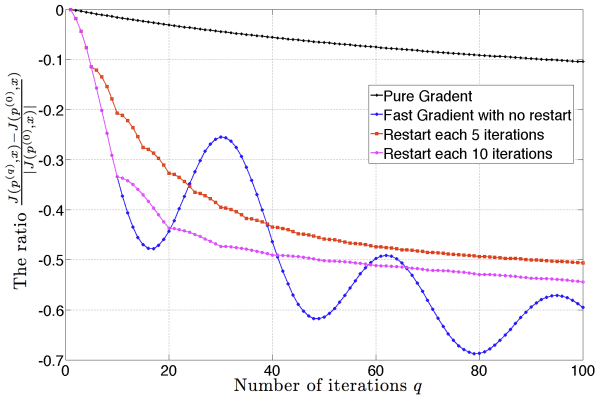

Fig. 1. Evolution of the relative decrease given by $\frac{J(p^{(q)},x) - J(p^{(0)},x)}{|J(p^{(0)},x)|}$ as a function of the number of iterations $q$ for pure gradient and fast gradient without restart or with two different values of the restart counter threshold $s_{max}$ used in Algorithm 3

---

**Algorithm 3** Fast gradient iterations $\bar{p} = \mathcal{S}^{(q)}(p, x)$ with constant restarting strategy

1: Initialization. $p^{(0)} \leftarrow p$, $r \leftarrow p$, $s \leftarrow 0$
2: **for** $i = 1, q$ **do**
3: $\quad s \leftarrow s + 1$
4: $\quad p^{(i)} \leftarrow P_C\big(r - \frac{1}{L}[\nabla J(r,x)]\big)$
5: $\quad r \leftarrow p^{(i)} + c(p^{(i)} - p^{(i-1)})$
6: $\quad$ **if** $(s = s_{max})$ **then** $r \leftarrow p^{(i)}$, $s = 0$ **End if**
7: **end for**
8: $\bar{p} \leftarrow p^{(q)}$

---

Note that in this new version, a restarting feature is introduced each $s_{max}$ iterations of the original fast algorithm in order to avoid too large moment due to increasing difference between $r$ and $p^{(i)}$.

Figure 1 shows a typical evolutions of the ratio:

$$\frac{J(p^{(q)},x) - J(p^{(0)},x)}{|J(p^{(0)},x)|}$$

for different configuration of Algorithm 3, namely: The pure gradient algorithm ($c = 0$), the fast gradient with no restart ($s_{max} = \infty$), the fast gradient with $s_{max} = 5$ and 8 respectively. These curves have been at a randomly chosen instant during the scenario depicted in the example studied in section V. Other random instant systematically gave the same kind of behavior.

The main message that can be derived from this observation is that provided that the restart technique is applied in the case of fast gradient ($c \neq 0$), the efficiency map is monotonic and a gradient descent can be safely used in the updating rule for the control updating period following the lines used in Algorithm 1.

## V. ILLUSTRATIVE EXAMPLE

Let us consider the constrained system consisting in a triple integrator:

$$\dot{x}_1 = x_2 \ ; \ \dot{x}_2 = x_3 \ ; \ \dot{x}_3 = u \ ; \ |u| \leq 1 \quad (23)$$

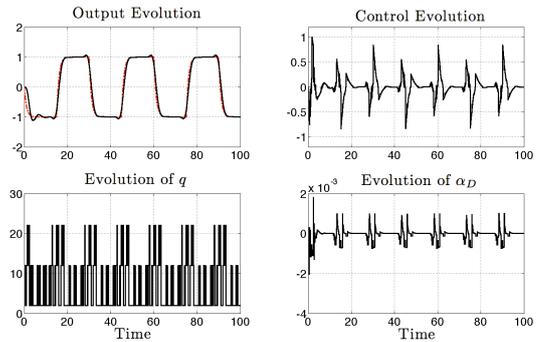

Fig. 2. Closed-loop evolution under the adaptive control updating period with initial $q = 2$, $\delta = 10$, $q_{max} = 100$. Prediction horizon $N = 200$.

and assume that fast gradient MPC is used to force this system to track some reference signal $y^{ref} = x_1^{ref}$ using the cost function:

$$J = \sum_{k=1}^{N} \|y(k) - y^{ref}(k)\|_Q^2 + \|u(k)\|_R^2 \quad (24)$$

with $Q = 100$ and $R = 1$ and a sampling period $\tau = 0.02$. The fast gradient is used with restart parameter $s_{max} = 8$. The number of iterations is limited to $q_{max} = 100$ and the step size $\delta = 10$ is used in the updating rule defined in step 6 of Algorithm 1.

Figures 2 and 3 show the behavior of the closed-loop using the proposed adaptive algorithm for two different initial values $q = 2$ and $q = 100$ respectively and using a prediction horizon length of $N = 200$ sampling periods. The results clearly show that the updating scheme converges towards the same pattern regardless of the initial choice of $q$. The performance of the closed-loop under the proposed adaptive strategy is to be compared to those depicted on Figures 4 and 5 where constant $q = 2$ and $q = 100$ are respectively applied to assess the efficiency of the proposed strategy to improve the convergence towards the ideal case.

Observing the behavior of the updating parameter $q$ on Figures 2 and 3 suggests that a constant $q = 20$ would be appropriate. This is checked on Figure 6 where indeed a constant $q = 20$ seems to give nice results. Nevertheless, the same constant value $q = 20$ is no more appropriate when a prediction horizon $N = 100$ is used as shown on Figure 7. Again, firing the updating mechanism and starting from $q = 20$ enables the new *optimal* updating parameter to be recovered leading to quasi-optimal performance level (Figure 8).

## VI. CONCLUSION

In this work a new updating mechanism for the number of iterations to be performed in fast gradient based NMPC implementation is proposed and assessed using a,n illustrative example. The method shows nice ability to automatically recover the optimal performance despite bad a priori knowledge on the optimal number of iterations. The updating mechanism uses cheap computation based on the behavior of the algorithm and the cost function in closed-loop.

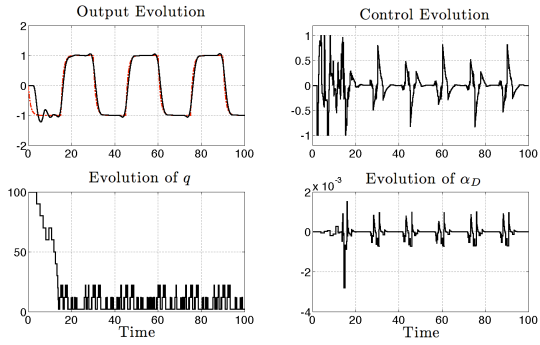

Fig. 3. Closed-loop evolution under the adaptive control updating period with initial $q = 100$, $\delta = 10$, $q_{max} = 100$. Prediction horizon $N = 200$

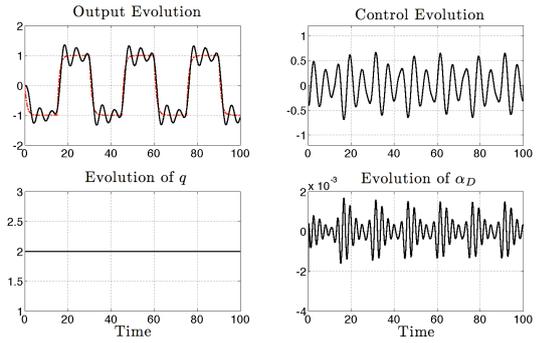

Fig. 4. Closed-loop evolution without the updating mechanism and using constant $q = 2$. Prediction horizon $N = 200$

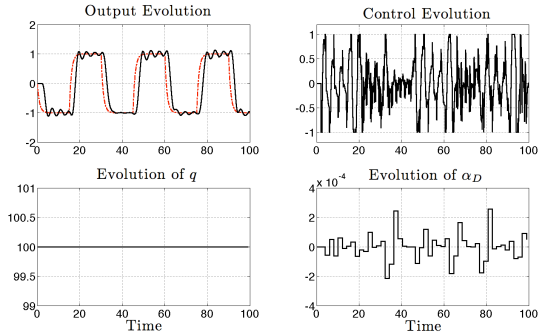

Fig. 5. Closed-loop evolution without the updating mechanism and using constant $q = 100$. Prediction horizon $N = 200$

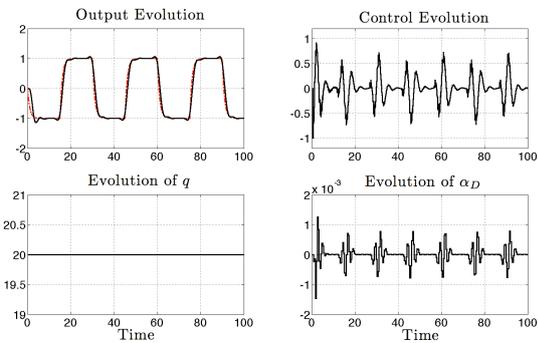

Fig. 6. Closed-loop evolution without the updating mechanism and using constant $q = 20$. Prediction horizon $N = 200$

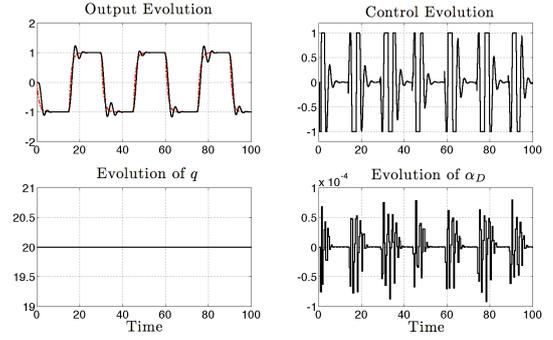

Fig. 7. Closed-loop evolution without the updating mechanism and using constant $q = 20$. Prediction horizon $N = 100$

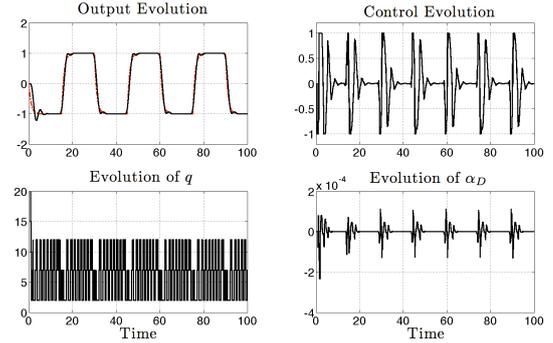

Fig. 8. Closed-loop evolution under the adaptive control updating period with initial $q = 20$, $\delta = 10$, $q_{max} = 100$. The restarting threshold $s_{max} = 8$ is used in the fast gradient descent. Prediction horizon $N = 100$

## REFERENCES

[1] M. Alamir. *Stabilization of nonlinear systems using receding-horizon control schemes: A parameterized approach for fast systems*. Springer-Verlag, 2006.
[2] M. Alamir. *A Framework for Monitoring Control Updating Period in Real-Time NMPC*, chapter In Assessement and Future Directions in Nonlinear Model Predictive Control. Lecture Notes in Control and Information Sciences, Springer-Verlag,, 2008.
[3] A. Bomporad and P. Patrinos. Simple and certifiable quadratic programming algorithms for embedded linear model predictive control. In *Proceeding of the IFAC Nonlinear Predictive Control Conference*, Noordwijkerhout, NL, 2012.
[4] C. N. Jones, A. Domahidi, M. Morari, S. Richter, F. Ullmann, and M. Zeilinger. Fast predictive control: Real-time computation and certification. In *Proceeding of the IFAC Nonlinear Predictive Control Conference*, Noordwijkerhout, NL, 2012.
[5] Y. Nesterov. A method of solving a convex programming problem with convergence rate o (1/k2). *Soviet Mathematics Doklady*, 27(2):372–376, 1983.
[6] Y. Nesterov. *Introductory lectures in convex optimization: a basic course*. Kluwer Academic Publishers, 2004.
[7] B. O'Donoghue and A. Candes. Adaptove restart for accelerated gradient schemes. arxiv:1204.3982. April 2012.
[8] P. Zometa, M. Kogel, T. Faulwasser, and R. Findeisen. Implementation aspects of model predictive control for embedded systems. In *Proceeding of the American Control Conference*, Washington, USA., 2012.